\documentclass[preprint,12pt]{elsarticle}

\usepackage{amssymb,amsmath}
\usepackage{amsthm}
\biboptions{sort&compress}
\newdefinition{rmk}{Remark}
\DeclareMathOperator{\tr}{tr}
\DeclareMathOperator{\diag}{diag}

\newcommand{\subfigimg}[3][,]{%
  \setbox1=\hbox{\includegraphics[#1]{#3}}
  \leavevmode\rlap{\usebox1}
  \rlap{\hspace*{0pt}\raisebox{\dimexpr\ht1-0.3\baselineskip}{\scriptsize{#2}}}
  \phantom{\usebox1}
}

\usepackage{makecell}
\DeclareMathOperator*{\argmax}{arg\,max}
\DeclareMathOperator*{\argmin}{arg\,min}

\newcommand{\inp}{\boldsymbol{x}}
\newcommand{\out}{\boldsymbol{y}}
\newcommand{\outzero}{\boldsymbol{\Delta y}}
\newcommand{\mout}{\boldsymbol{z}}
\renewcommand{\d}{\mathrm{d}}
\newcommand{\bayesequal}{\mout=\bar{\out}}

\journal{Journal}

\begin{document}

\begin{frontmatter}

\title{A Bayesian constitutive model selection framework for biaxial mechanical testing of planar soft tissues: application to porcine aortic valves}

\author[gcec]{Ankush Aggarwal\corref{cor1}}

\affiliation[gcec]{organization={Glasgow Computational Engineering Centre, James Watt School of Engineering},
            addressline={University of Glasgow}, 
            city={Glasgow},
            postcode={G12 8LT}, 
            state={Scotland},
            country={United Kingdom}}

\author[ou]{Luke T. Hudson}
\author[ou]{Devin W. Laurence}
\author[ou]{Chung-Hao Lee}

\affiliation[ou]{organization={Biomechanics and Biomaterials Design Laboratory, School of Aerospace and Mechanical Engineering},
            addressline={The University of Oklahoma}, 
            city={Norman},
            postcode={73019}, 
            state={OK},
            country={United States of America}}

\author[swansea]{Sanjay Pant}

\affiliation[swansea]{organization={Faculty of Science and Engineering},
            addressline={Swansea University}, 
            city={Swansea},
            postcode={SA1 8EN}, 
            state={Wales},
            country={United Kingdom}}

\cortext[cor1]{Correspondence to ankush.aggarwal@glasgow.ac.uk}

\begin{abstract}
{A variety of constitutive models have been developed for soft tissue mechanics. However, there is no established criterion to select a suitable model for a specific application. Although the model that best fits the experimental data can be deemed the most suitable model, this practice often can be insufficient given the inter-sample variability of experimental observations. Herein, we present a Bayesian approach to calculate the relative probabilities of constitutive models based on biaxial mechanical testing of tissue samples. 46 samples of porcine aortic valve tissue were tested using a biaxial stretching setup. For each sample, seven ratios of stresses along and perpendicular to the fiber direction were applied. The probabilities of eight invariant-based constitutive models were calculated based on the experimental data using the proposed model selection framework. The calculated probabilities showed that, out of the considered models and based on the information available through the utilized experimental dataset, the May--Newman model was the most probable model for the porcine aortic valve data. When the samples were grouped into different cusp types, the May--Newman model remained the most probable for the left- and right-coronary cusps, whereas for non-coronary cusps two models were found to be equally probable: the Lee--Sacks model and the May--Newman model. This difference between cusp types was found to be associated with the first principal component analysis (PCA) mode, where this mode's amplitudes of the non-coronary and right-coronary cusps were found to be significantly different. Our results show that a PCA-based statistical model can capture significant variations in the mechanical properties of soft tissues. The presented framework is applicable to any tissue type, and has the potential to provide a structured and rational way of making simulations \emph{population-based}.}
\end{abstract}



\end{frontmatter}

\section{Introduction}
Soft tissues exhibit a complex stress-strain behavior, including nonlinearity and anisotropy, that varies not only across tissue types, but also from sample to sample. Decades of research into the biomechanics of soft tissues has shed important light on their behavior and role in many physiological systems, such as vascular, lungs and ligaments. However, there are still open challenges that need further investigations. One of these challenges is modeling the biomechanical behavior of soft tissues reliably and with high fidelity. This challenge remains an active area of research. 

Numerous constitutive models have been developed to describe the stress-strain behavior of soft tissues \cite{maurel1998}. These models range from purely phenomenological to multi-scale ones that incorporate detailed microstructural information. Some of the more commonly adopted models can be categorized into Fung-type \cite{fung1972stress}, invariant-based \cite{doi:10.1098/rsif.2005.0073}, and structural models \cite{lanir1983constitutive, billiar2000biaxial}. These model categories have individual pros and cons. For example, the Fung-type models, without additional treatment, do not satisfy frame invariance \cite{ATESHIAN2009781, sun2005finite}, while the structural models are computationally too expensive to be employed in finite element simulations of realistic biological systems.

Even within each category, there are a large number of available models that can be challenging to differentiate. It is often unclear which model is most suitable for a given problem or situation, thus making selection of a particular model challenging. While the model that best fits a tissue's ex-vivo response may be considered an ``optimal'' choice, different definitions of ``best fit'' can lead to different results. For example, how one prescribes relative weights to different experimental protocols performed on a tissue sample can have an effect on the fit. This becomes a unique challenge when none of the models fit all the experiments simultaneously, leading to a trade-off when performing the fitting. Moreover, considering the inter-sample variability in many biological systems, there is no guarantee that a model that fits the data for one sample will also be representative of the data for another sample of the same tissue type.  Nevertheless, it is reasonable to expect that a chosen model should be able to represent several (ideally, all) samples, not just one. 

The focus of the present study is on the problem of choosing a model for soft tissues, which is termed as ``model selection''. In general, model selection is a non-trivial problem, and several approaches have been proposed in the literature \cite{doi:10.1080/07474939208800232}, such as a Bayesian framework \cite{FARRELL2015189}, techniques based on cross-validation \cite{10.1214/09-SS054} and those based on information criteria \cite{10.1093/biomet/83.4.875}. However, these techniques are only starting to be used in the field of tissue biomechanics \cite{doi:10.1098/rsif.2020.0886,doi:10.1142/S0218202513500103, MADIREDDY2015102}. This is partly because model selection becomes all the more challenging due to the nonlinearities of constitutive models, the high dimension of the measurement space, and the subtle variations in how experiments are conducted.

An effective and widely used experimental method for biomechanical characterization of soft tissues is biaxial testing, which has been applied to various types of thin tissues \cite{10.1115/1.2835086,billiar2000biaxial,zhang2015generalized, HUMPHREY198759, VITO1980947, 10.1115/1.2894887, YIN1987577}. With established testing setups and fast acquisition commercial systems, it is now possible to collect biaxial test data on a large number of samples and employ advanced techniques from data science (e.g., machine learning) for solving unresolved issues. Thus, the goal of this study is to formulate a Bayesian framework for model selection that can be applied to data from planar biaxial mechanical testing. Herein, model selection is posed as a problem of \emph{selecting a model that has the highest probability given the experimental data}. Importantly, the framework is designed to account for the inter-sample variability and experimental noise within a Bayesian setting.

To demonstrate the proposed model selection framework, we apply it to aortic valve (AV) tissue, which is clinically important for healthy functioning of the heart. The AV is made up of three semilunar cusps: left coronary cusp (LCC), non-coronary cusp (NCC), and right coronary cusp (RCC). While biomechanics of the AV tissue has been studied extensively in the literature \cite{billiar2000biaxial, WU201823, 10.1002/jbm.a.34099, 10.1016/j.ejcts.2004.05.043, ECKERT20134653, doi:10.1089/ten.2006.0279, HASAN20141949, SAUREN1983327, balachandran2011hemodynamics, anssari2011combined},  there is no consensus yet regarding its most appropriate constitutive model \cite{10.1115/1.3127261, 10.1115/1.1894373, auricchio2012comparison, sun2005finite}. Further, the three cusp types also pose an interesting question: can the same model be used to represent all three AV cusps or a different model is required for each cusp type? 

The proposed framework aims to be general and applicable to all tissue types, while also providing a unique insight into the biomechanics of AV tissue. This article is organised as follows. The experimental, theoretical, and computational methods are described in Section~2. Then, the results using the proposed framework for AV tissue are presented in Section~3. Finally, the implications and potential uses of the proposed framework are discussed in Section~4.

\section{Methods}

\subsection{Data generation and pre-processing}
A pre-requisite for the proposed framework is the availability of data from a sufficient number of samples to generate a statistical model. In this subsection, the details of the experimental setup used to generate the data and the techniques used for pre-processing of the data are presented. The experimental data used in this study is the same as that reported in a previous study \cite{HUDSON2022104907}, and its experimental procedure is summarized next, followed by the details of data pre-processing required for the proposed framework to work.

\subsubsection{Tissue preparation}
Eighteen porcine hearts (80--140 kg of weight, 1--1.5 years of age) were obtained from a USDA-approved abattoir (Chickasha Meat Company, Chickasha, OK). Each heart was dissected, and the three AV cusps (LCC, NCC, and RCC) were extracted from the aortas. The cusps were then briefly stored at $-20^\circ$C prior to mechanics testing within 6--12 hours. Prior to biaxial testing, the excised AV specimens were thawed in an in-house phosphate-buffered saline (PBS) solution at room temperature. Once thawed, the belly region of the tissue was dissected from the cusp, and thickness measurements were made using a non-contact laser displacement sensor (Keyence IL-030, Itaska, IL) at three different locations of each cusp specimen to determine the average tissue thickness. 

\subsubsection{Biaxial mechanical testing protocols}

For biaxial testing, the tissue specimens were mounted to a commercial biaxial testing system (BioTester, CellScale, Canada, 1.5~N load cells) via BioRake tines, resulting in an effective testing region of $6.5\times6.5$ mm. During mounting, the tissue's circumferential and radial directions were aligned with the $x$- and $y$-directions of the biaxial testing system, respectively. Four glass beads (with a diameter of $300-500$ {\textmu}m) were placed on the center region of each specimen to serve as fiducial markers for quantifying the in-plane strains.

\begin{figure}[h!]
\centering
\includegraphics[width=0.8\columnwidth]{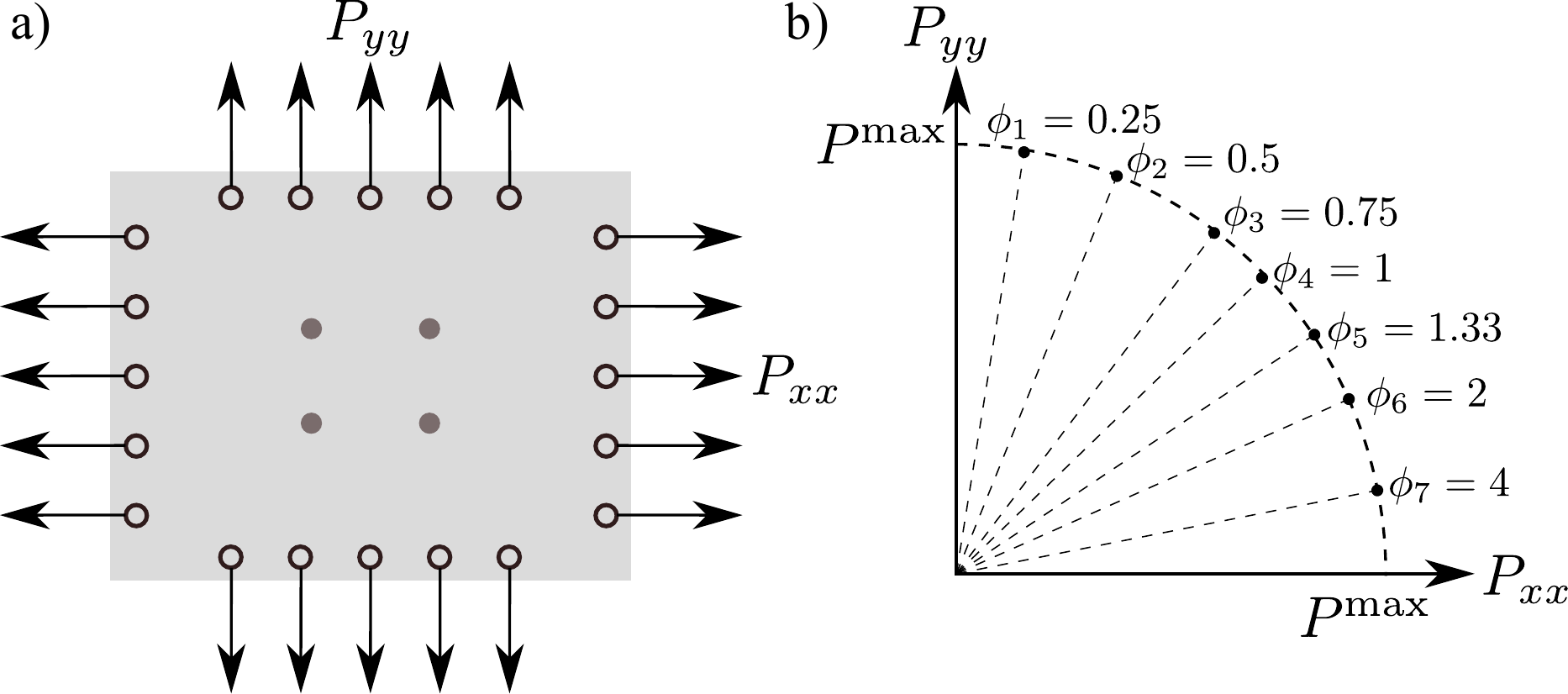}
\caption{(a) A schematic of the biaxial testing with the circumferential (fiber) direction of the tissue was aligned with the $x$ axis and the radial (cross-fiber) direction of the tissue was aligned with the $y$ axis. (b) The different loading paths in the stress space with seven different loading ratios $\phi_r \in \left\{ 0.25, 0.5, 0.75, 1, 1.333, 2, 4 \right\}$.\label{schematic}}
\end{figure}

For testing, the specimen were submerged in a 32$^\circ$C PBS bath during the testing. The force readings from the load cells and CCD camera images were recorded at 15 Hz throughout the test. The biaxial loading rates were restricted to $<3.32$\%/sec to be within the quasi-static loading range ($<12$\%/sec) to minimize any potential effects of strain rate on the results. At any point, if $f_x$ and $f_y$ were the forces applied in the $x$- and $y$-directions, respectively, the measured normal stresses were calculated as $P_{xx} = {f_x}/{tL_y}$ and $P_{yy} = {f_y}/{tL_x}$, where $L_x$ and $L_y$ are the effective dimensions of the sample and $t$ is the average measured tissue thickness in the unloaded configuration (Fig.~\ref{schematic}a). The deformation gradient $\mathbf{F}$ was quantified using bi-linear interpolation of the bead positions \cite{HUDSON2022104907}, and the right Cauchy--Green deformation tensor was calculated as $\mathbf{C}=\mathbf{F}^\top\mathbf{F}$ (here $(\cdot)^\top$ denotes the matrix transpose). Since the tissue's fiber orientation was aligned with the biaxial testing direction in the experimental setting, the off-diagonal terms in the deformation tensor were assumed to be small, effectively neglecting any shear deformation. The stretches along the two axes were calculated as $\lambda_x=\sqrt{C_{xx}}$ and $\lambda_y=\sqrt{C_{yy}}$, where $C_{xx}$ and $C_{yy}$ are the two diagonal components of $\mathbf{C}$. The stretch in the tissue's thickness direction was calculated using the incompressibility constraint, i.e., $\lambda_z={1}/{\lambda_x\lambda_y}$. 

A preconditioning protocol, consisting of six loading/unloading cycles at a target first Piola-Kirchhoff (PK) peak stress of $P = 240$ kPa, was first applied to restore the tissue to its in-vivo biomechanical configuration. The preconditioning protocols were followed by the actual testing protocols. Each testing protocol was defined as recording stresses and stretches along a loading path in the $P_{xx}$-$P_{yy}$ space starting at zero-stress state and ending at a target maximum stress $[P_{xx}^{r,\text{max}},P_{yy}^{r,\text{max}}]$. The target maximum stress state for a protocol $r$ had an associated ratio, $\phi_r={P_{xx}^{r,\text{max}}}/{P_{yy}^{r,\text{max}}}$ and target stress magnitude $P^{\text{max}} = \sqrt{\left(P_{xx}^{r,\text{max}}\right)^2+\left(P_{yy}^{r,\text{max}}\right)^2}$ (Fig.~\ref{schematic}b), with $r=1,\dots,R$. Target stress magnitude was kept approximately constant across all samples and protocols, while the target ratio was varied between protocols, so that $[P_{xx}^{r,\text{max}},P_{yy}^{r,\text{max}}] = \dfrac{P^{\text{max}}}{\sqrt{1+\phi_r^2}} [\phi_r,1]$. The corresponding maximum stretch for each protocol was pre-determined and then stretches were increased linearly from the reference state ($\lambda_x=\lambda_y=1$) to reach the maximum stretch (and therefore the maximum stress) state. For protocol $r$, $m_r$ points were recorded, and, therefore, for each sample,  $\sum_{r=1}^{R} 2m_r$ stresses and $\sum_{r=1}^{R} 2m_r$ stretches were recorded. 

\subsubsection{Data collection}
$N=46$ samples of aortic valve tissue were tested (15 LCC, 15 RCC, and 16 NCC). $R=7$ target ratios were used for each tissue sample, with $\phi_r \in \left\{ 0.25, 0.5, 0.75, 1, 1.333, 2, 4 \right\}$. Each protocol was repeated for three loading/unloading cycles, and the measurements from the final loading cycle were used. In practice, the actual maximum stress magnitude $P^{\text{max}}$ and the actual stress ratios $\phi_r$ varied slightly from the target values. More importantly, the number of points along the curve $m_r$ varied from sample to sample. As a result each sample had different number of measurements. In order to create a uniform number of measurements across samples, an interpolation of the experimentally measured stretch-stress curves was required, which is described next. 

\subsubsection{Interpolation and smoothing}
To standardize the measurements for all samples with the same range of applied stresses and the same number of measurement points, an interpolation was necessary. An appropriate interpolation function was required that provided a good fit to the full range of stress-stretch curves. After testing various options, the following one-dimensional function based on implicit elasticity proposed by Freed and Rajagopal \cite{freed2016promising} was used
\begin{equation}
\varepsilon = \varepsilon^C + \varepsilon^E 
            = \frac{\sigma}{E^C} + \frac{1}{\beta}\left( 1 - \frac{1}{\left( 1 + (\beta-1)\sigma/E^E \right)^{\beta/(\beta-1)}} \right),\label{freed-func}
\end{equation}
where $\varepsilon:=\lambda_{x}-1$ and $\sigma:=P_{xx}$ for curves along the fiber direction, and $\varepsilon:=\lambda_{y}-1$ and $\sigma:=P_{yy}$ for the cross-fiber direction. The above function has three parameters, $\left\{\beta,E^E,E^C\right\}$, which were determined by fitting Eq.~\eqref{freed-func} to each experimental stress-stretch curve. Since the one-dimensional function is based on implicit elasticity theory, it also helped avoid any non-physical oscillations in the interpolated data. 

After fitting the above function to each experimental stress-stretch curve, an interpolated (or extrapolated) and smoothed version of the dataset was produced, with each curve having $\bar{m}=100$ points and reaching a maximum stress magnitude of $\left< P^{\text{max}}\right>$ (here $\left<\cdot\right>$ denotes the mean operator over all tissue samples). 
Thus, after this step, each sample $I$ had the same input 
\begin{equation}
\inp=\bigcup_{r=1}^R \left\{P_{xx}^{r,\text{max}},P_{yy}^{r,\text{max}} \right\}.\label{input}
\end{equation}
The outputs included measured stresses, denoted as a vector $\boldsymbol{\sigma}^{(I)}\in\mathbb{R}^{2R\bar{m}}$, and the same number of stretches, denoted as a vector $\boldsymbol{\lambda}^{(I)}\in\mathbb{R}^{2R\bar{m}}$. Since the stretches varied linearly for each protocol, the stretch vector could be represented simply in terms of the maximum stretches for each protocol, $\boldsymbol{\lambda}^{\text{max},(I)}\in\mathbb{R}^{2R}$. The combination of normalized stresses and maximum stretches for each sample was represented with a combined output vector  
\begin{equation}
\out^{(I)}:=\frac{\boldsymbol{\sigma}^{(I)}}{\left< P^{\text{max}}\right>} \cup \boldsymbol{\lambda}^{\text{max},(I)},\label{output}
\end{equation}
and this combined output vector $\out^{(I)}\in\mathbb{R}^{2R(\bar{m}+1)}$.

\subsubsection{Principal component analysis}
After interpolation, we had the same number of measurements for all $N$ samples, $\out^{(I)}$, $I=1,\dots,N$. From these measurements, a statistical distribution of the measured output was sought. A fundamental statistical distribution is the multivariate Gaussian distribution, which requires estimation of the mean vector and covariance matrix. However, because of the high-dimensionality of the output space, directly estimating its covariance matrix would have required a prohibitively large number of samples. Therefore, a reduction in dimensionality was first achieved via principal component analysis (PCA) as follows. 

First, the mean output was calculated as
\begin{equation}
\bar{\out} = \frac{1}{N}\sum_{I=1}^N \out^{(I)},
\end{equation}
and a zero-mean output vector for each sample was calculated as
\begin{equation}
\outzero^{(I)} = \out^{(I)}-\bar{\out}.
\end{equation}
All the zero-mean output vectors were written in a matrix form $\mathbf{Z}$, where the $I^\mathrm{th}$ row is $\outzero^{(I)}$\footnote{Remark: A direct estimation of the covariance matrix would be $\mathbf{\Sigma} = \mathbf{Z}^\top\mathbf{Z}$. However, since $N\ll 2R(\bar{m}+1)$, this estimate of the covariance matrix would be extremely ill-conditioned and not usable for building a statistical model. Thus, a PCA was required to resolve this issue.}. Next, a singular value decomposition of $\mathbf{Z}$ was performed as
\begin{equation}
\mathbf{Z} = \mathbf{U}\mathbf{S}\mathbf{V}^{\mathsf{H}},
\end{equation}
where $(\cdot)^\mathsf{H}$ denotes the conjugate transpose of a matrix, $\mathbf{S}$ is a diagonal matrix with singular values $s_\alpha$ (equal to the square root of the eigenvalues of $\mathbf{Z}^{\mathsf{H}}\mathbf{Z}$), and rows of $\mathbf{V}^{\mathsf{H}}$, $\boldsymbol{v}_\alpha$, are the corresponding unitary eigenvectors (also called principal modes) of $\mathbf{Z}^{\mathsf{H}}\mathbf{Z}$. The singular values and vectors pairs were written as $(s_\alpha,\boldsymbol{v}_\alpha)$, with $\alpha=1,\dots,N$. The PCA mode amplitudes of each data set were calculated as 
\begin{equation}
a_\alpha^{(I)} = \outzero^{(I)} \cdot \boldsymbol{v}_{\alpha}.
\end{equation}
Therefore, the reconstructed measurements from first $M$ principal modes were
\begin{equation}
\tilde{\out}^{(I)} = \sum_{\alpha=1}^M a_\alpha^{(I)} \boldsymbol{v}_{\alpha} + \bar{\out}.
\end{equation}
Using the unitary property of $\mathbf{U}$, it is easy to see that the singular values $s_\alpha$ also represent the standard deviation of the modal amplitudes $a_\alpha^{(I)}$. Thus, keeping the first $M$ principal modes, the experimental data was represented as a statistical model\footnote{Remark: Choosing a statistical model here can also be considered as a problem of model selection. Since the mode amplitudes are scalars and independent (because of PCA), this is an easier problem. For simplicity, a normal distribution was chosen for the modal amplitudes. However, if enough samples are available, it is possible to select more appropriate distributions for each PCA mode.}
\begin{equation}
\boldsymbol{Y} = \bar{\out} + \sum_{\alpha=1}^{M} \mathcal{N}(0,s_\alpha^2) \boldsymbol{v}_\alpha +  \left(\mathbf{I}-\sum_{\alpha=1}^M \boldsymbol{v}_\alpha \otimes \boldsymbol{v}_\alpha \right) \boldsymbol{\epsilon},\label{statistical-model}
\end{equation}
where $\boldsymbol{\epsilon}$ is a random vector with norm $\| \boldsymbol{\epsilon}\| \sim \mathcal{N}(0,\sigma^2+\sigma_n^2)$. The variance of this random vector was related to the fact that $\alpha>M$ modes were truncated: 
\begin{equation}
\sigma^2 = \frac{1}{N} \sum_{I=1}^{N} \|\out^{(I)} - \tilde{\out}^{(I)}\|^2,
\end{equation}
whereas the measurement noise variance $\sigma_n^2$ was calculated from the interpolation error.

\begin{table*}[ht]
\resizebox{\textwidth}{!}{\begin{tabular}{lll}
\hline
\bf{Model} & \bf{Strain energy density function} & \bf{Parameters} \\
\hline

GOH & $\Psi = \frac{\mu}{2}(I_1-3) + \frac{k_1}{2k_2}\left[\exp(k_2(\kappa I_1+(1-3\kappa)I_4-1)^2)-1\right]$ & $\theta = \left\{\mu,k_1,k_2,\kappa\right\}$ \\[3pt]

HGO & $\Psi =  \frac{\mu}{2}(I_1-3) + \frac{k_1}{2k_2}\left[\exp(k_2(I_4-1)^2)-1\right]$ & $\theta = \left\{\mu,k_1,k_2\right\}$ \\[3pt]

HGO2 & $\Psi = \frac{k_1}{k_2}\left[\exp(k_2(I_1-3))-1\right] + \frac{k_3}{2k_4}\left[\exp(k_4(I_4-1)^2)-1\right]$ & $\theta = \left\{k_1,k_2,k_3,k_4\right\}$ \\[3pt]

Holzapfel & $\Psi = \frac{\mu}{2}(I_1-3) +  \frac{k_1}{2k_2}\left[ \exp(k_2(\kappa(I_1-3)^2+(1-\kappa)(I_4-1)^2))-1\right]$ & $\theta = \left\{\mu,k_1,k_2,\kappa\right\}$ \\[3pt]

HY & $\Psi = \frac{k_1}{k_2}\left[\exp(k_2(I_1-3))-1\right] +  \frac{k_3}{k_4}\left[\exp(k_4(\sqrt{I_4}-1)^2)-1\right]$ & $\theta = \left\{k_1,k_2,k_3,k_4\right\}$ \\[3pt]

LS & $\Psi = \frac{\mu}{2}(I_1-3) + \frac{k_1}{2k^*}\left[\kappa\exp(k_2(I_1-3)^2) + (1-\kappa)\exp(k_3(I_4-1)^2)-1\right]$  & $\theta = \left\{\mu,k_1,k_2,k_3,\kappa\right\}$ \\[3pt]

MN & $\Psi = \frac{\mu}{2}(I_1-3) + \frac{k_1}{k_2+k_3}\left[\exp(k_2(I_1-3)^2+k_3(\sqrt{I_4}-1)^4)-1\right]$ & $\theta = \left\{k_1,k_2,k_3,\mu\right\}$ \\[3pt]

Yeoh & $\Psi = \sum_{i=1}^3 c_i (I_1-3)^i $ & $\theta = \left\{c_1,c_2,c_3\right\}$ \\[3pt]
\hline
\end{tabular}}
\caption{List of models considered with their strain energy density functions and the associated parameters. For LS model, $k^*:=\kappa k_2+(1-\kappa)k_3$ is used for brevity.}
\label{model-list}
\end{table*}

\subsection{Hyperelastic constitutive models}
The main question this study aims to address is, ``which model should be selected given the data from $N$ samples described above?''. In order to proceed, eight hyperelastic constitutive models that have been developed for soft tissues were pre-selected. The choice, although not an exhaustive list, covers several invariant-based models that can be difficult to differentiate. The following models were considered in this study: (i) an isotropic model by Yeoh for rubber elasticity \cite{yeoh-model}; (ii) the Lee--Sacks (LS) model for the mitral valve leaflet tissue \cite{LEE20142055}; (iii) the May--Newman (MN) model with another form proposed for the mitral valve tissue \cite{may1998constitutive}; (iv and v) two variants of a model proposed by Holzapfel, Gasser, and Ogden for arterial tissue with an additive split of isotropic and anisotropic components \cite{holzapfel2000new} (HGO with linear isotropic term and HGO2 with an exponential isotropic term); (vi) Holzapfel model proposed for coronary arteries \cite{doi:10.1152/ajpheart.00934.2004}; (vii) another model proposed by Gasser, Ogden and Holzapfel (GOH) for coronary arteries \cite{doi:10.1098/rsif.2005.0073}, and (viii) Humphrey--Yin (HY) model developed for myocardium \cite{HUMPHREY1987563}. Some theoretical limitations have been reported for these models in the literature, however the shortlist was made based on their common use in practice.

Hyperelastic models define a strain energy density function (SEDF) $\Psi$. The SEDFs and corresponding parameters $\theta$ of all the eight models in alphabetical order are summarized in Table~\ref{model-list}. From the SEDF, the first PK stress is derived as \cite{holzapfel-book}
\begin{equation}
\mathbf{P} = \frac{\partial \Psi}{\partial \mathbf{F}} - p \mathbf{F}^{-\top},
\end{equation}
where $p$ is the hydrostatic pressure to enforce incompressibility. Based on the applied deformation in the biaxial setup, we can determine the deformation gradient $\mathbf{F} = \diag[\lambda_x, \lambda_y,{1}/{\lambda_x\lambda_y}]$. The models considered are functions of the first invariant $I_1=\tr(\mathbf{C})$ and the fourth invariant $I_4=\mathbf{N}\cdot\mathbf{C}\mathbf{N}$, where $\mathbf{N}$ is the fiber direction and approximated to be along the $x$-axis. Thus, $\dfrac{\partial \Psi}{\partial \mathbf{C}} = \dfrac{\partial \Psi}{\partial I_1}\mathbf{I} + \dfrac{\partial \Psi}{\partial I_4} \mathbf{N}\otimes\mathbf{N}$, and the hydrostatic pressure $p$ is analytically derived by equating the normal stress along tissue's thickness, $P_{zz}=0$ \cite{FAN20142043, KIENDL2015280}. 

Thus, given a model for SEDF $\Psi$, the resulting stresses $P_{xx}$ and $P_{yy}$ can be obtained from stretches $\lambda_x$ and $\lambda_y$. However, since the experiments were performed to target stresses $[P_{xx}^{r,\text{max}},P_{yy}^{r,\text{max}}]$, the inputs to the model were the stresses instead. From these maximum target stresses, the maximum target stretches were computed iteratively using a modified Powell method implementation in SciPy \cite{powell-method}. Once the target stretches were found, $\bar{m}$ equi-spaced stretch increments were applied to find the resulting stresses. The resulting stresses were then normalized by $\left< P^{\text{max}}\right>$ and combined with the maximum target stretches (as per Eq.~\ref{output}) to obtain the model output vector, denoted as $\mout$ which is a function of the chosen model and associated parameter values.

Note that, for some models, their parameterization was slightly altered from the original versions to make the parameters comparable to other models. Moreover, all the considered models have similar numbers (3 to 5) of parameters. For any additional models to be considered, the framework can simply be applied to the new model and the results compared with those presented here. 


\subsection{Proposed framework: Bayesian model selection}
In this subsection, the framework for model selection is described. If $\mathcal{K}$ models are considered with output $\mout=\mathcal{M_I}(\inp,\theta)$, where $\mathcal{M_I}$ represents the $\mathcal{I}^\mathrm{th}$ model with associated parameters $\theta$ and $\mathcal{I}=1,\dots,\mathcal{K}$, these were compared to the statistical model of the measurements $\boldsymbol{Y}$ in Eq.~\eqref{statistical-model} as follows. From the Bayes' theorem (see \ref{bayes-notation} for the preliminaries), we have:
\begin{equation}
p(\theta \mid \bayesequal ,\mathcal{M_I}) = \frac{p(\bayesequal \mid \theta, \mathcal{M_I})p(\theta \mid \mathcal{M_I})}{p(\bayesequal \mid \mathcal{M_I})},\label{bayes1}
\end{equation}
where the denominator on the right-hand side is an integral of the numerator, i.e.,
\begin{equation}
p(\bayesequal \mid \mathcal{M_I}) = \int\limits_\theta p(\bayesequal \mid \theta, \mathcal{M_I})p(\theta \mid \mathcal{M_I}) \, \d\theta .\label{main-integral}
\end{equation}
The above integral balances model complexity and quality of fit by rewarding the goodness of fit while penalising models with parameters that do not contribute to the goodness of fit. Applying the Bayes' theorem once again, we arrived at the probability of model $\mathcal{I}$ given the measurements
\begin{equation}
p(\mathcal{M_I} \mid  \bayesequal) = \frac{p(\bayesequal \mid \mathcal{M_I}) p(\mathcal{M_I})}{p(\bayesequal)},
\end{equation}
where the denominator is the summation over the numerator, i.e.,
\begin{equation}
p(\bayesequal) = \sum\limits_{\mathcal{I}=1}^{\mathcal{K}} p(\bayesequal \mid \mathcal{M_I}) p(\mathcal{M_I}).
\end{equation}
This approach, also known as the Bayes factor, has been proposed to compare any two models \cite{doi:10.1080/01621459.1995.10476572} and is being recently used in mechanics \cite{GIROLAMI2021113533,fitt2019uncertainty,MADIREDDY2015102}. Thus, to evaluate the model probabilities $p(\mathcal{M_I} \mid  \bayesequal)$, Eq.~\eqref{main-integral} has to be evaluated with specified or assumed prior probabilities of models $p(\mathcal{M_I})$. In the absence of any prior knowledge or preference, equal prior probabilities of the models were used, i.e., $p(\mathcal{M}_\mathcal{I})=1/\mathcal{K}$ for all $\mathcal{I}$. The integral in Eq.~\eqref{main-integral} was computed using Monte Carlo integration, as described next.

\begin{table*}[h!]
\resizebox{\textwidth}{!}{\begin{tabular}{llllll}
\hline
\bf{Model} & {\bf$\bar{\theta}_1$} (in kPa) & {\bf$\bar{\theta}_2$} (in kPa) & \bf{$\bar{\theta}_3$} & \bf{$\bar{\theta}_4$} & \bf{$\bar{\theta}_5$} \\
\hline

GOH & $\mu=3.68$  & $k_1=30.00$ & $k_2=8.93$ & $\kappa=0.30$ & -- \\

HGO & $\mu=28.48$  & $k_1=1.86$ & $k_2=5.68$ & --  & -- \\

HGO2 & $k_1=0.25$  & $k_3=0.64$ & $k_2=3.55$  & $k_4=0.25$ & -- \\

Holzapfel & $\mu=4.14$  & $k_1=4.89$ & $k_2=1/86$ & $\kappa=0.51$ & -- \\

HY & $k_1=0.25$ & $k_3=0.92$  & $k_2=3.54$ & $k_4=44.92$ & -- \\

LS & $\mu=5.3$  & $k_1=2.65$ & $k_2=1.21$ & $k_3=7.14$ &  $\kappa=0.96$ \\

MN & $\mu=4.21$ &  $k_1=57.45$  & $k_2=0.93$ & $k_3=36.08$ & -- \\

Yeoh & $c_1=0$  & $c_2=0$ & $c_3=6.8$ kPa & -- & -- \\

\end{tabular}}
\caption{Model parameters values $\bar{\theta}_i$ corresponding to the best classical fit to the mean response}
\label{fit-params}
\end{table*}

\begin{table*}[h!]
\resizebox{\textwidth}{!}{\begin{tabular}{llllll}
\hline
\bf{Model} & {\bf$\theta_1$ prior} (in kPa) & {\bf$\theta_2$ prior} (in kPa) & \bf{$\theta_3$ prior} & \bf{$\theta_4$ prior} & \bf{$\theta_5$ prior} \\
\hline

GOH  & $\mu \in[ 0.37 , 46.83 ]$ & $k_1 \in[ 3.00 , 310.00 ]$ & $k_2 \in[ 0.89 , 99.30 ]$ & $k_3 \in[ 0 , 1/3 ]$ &  \\

HGO  & $\mu \in[ 2.85 , 294.85 ]$ & $k_1 \in[ 0.19 , 28.57 ]$ & $k_2 \in[ 0.57 , 66.83 ]$ & -- & --  \\

HGO2 & $k_1 \in[ 0.02 , 12.47 ]$ & $k_3 \in[ 0.06 , 16.36 ]$ & $k_2 \in[ 0.36 , 45.54 ]$ & $k_4 \in[ 0.70 , 79.60 ]$ & -- \\

Holzapfel & $\mu \in[ 0.41 , 51.43 ]$ & $k_1 \in[ 0.49 , 58.94 ]$ & $k_2 \in[ 0.19 , 28.62 ]$ & $\kappa \in[ 0 , 1 ]$ & -- \\

HY & $k_1 \in[ 0.03 , 12.51 ]$ & $k_3 \in[ 0.09 , 19.17 ]$ & $k_4 \in[ 4.49 , 459.20 ]$ & $k_2 \in[ 0.35 , 45.43 ]$ & -- \\

LS  & $\mu \in[ 0.53 , 63.00 ]$ & $k_1 \in[ 0.26 , 36.47 ]$ & $k_2 \in[ 0.12 , 22.10 ]$ & $k_3 \in[ 0.71 , 81.45 ]$ & $\kappa \in[ 0 , 1 ]$ \\

MN  & $\mu \in[ 0.42 , 52.08 ]$ & $k_1 \in[ 5.75 , 584.50 ]$ & $k_2 \in[ 0.09 , 19.31 ]$ & $k_3 \in[ 3.61 , 370.77 ]$ & -- \\

Yeoh & $c_1 \in[ 0.00 , 10.00 ]$ & $c_2 \in[ 0.00 , 10.00 ]$ & $c_3 \in[ 0.68 , 78.01 ]$ kPa & -- & -- \\

\end{tabular}}
\caption{Prior distributions of the model parameters were assumed to be uniform in the following ranges (around the best classical fit from Table~\ref{fit-params})}
\label{model-priors2}
\end{table*}
\subsubsection{Monte Carlo integration}
The integral in Eq.~\eqref{main-integral} can be high-dimensional with a large or, possibly, infinite domain. Thus, Monte Carlo integration was used to approximate this integral \cite{MOROKOFF1995218}, i.e.,
\begin{equation}
p(\bayesequal \mid \mathcal{M_I}) \approx \dfrac{1}{S} \sum\limits_{s=1}^S p(\bayesequal \mid \theta^s, \mathcal{M_I}), \label{MC}
\end{equation}
where $\theta^s$, $s=1,\dots,S$, are samples from the prior distribution of model parameters $p(\theta \mid \mathcal{M_I})$. The useful property of Monte Carlo integration is that the approximation error converges $\sim \dfrac{1}{\sqrt{S}}$ independently of the dimension of the parameter space. Moreover, it is trivial to implement and parallelize. Lastly, the prior distribution can be sampled randomly or quasi-randomly, with the latter giving faster convergence in practice \cite{MOROKOFF1995218}. Therefore, a Sobol sequence \cite{10.1145/641876.641879,doi:10.1137/070709359} was used to generate $S=2^{15}$ samples from the prior distributions of parameters of each model, $p(\theta \mid \mathcal{M_I})$.

\subsubsection{Calculating the likelihood}
In Eq.~\eqref{MC}, it is required to calculate the likelihood function in the RHS. This is computed from the statistical model presented in Eq.~\eqref{statistical-model}. That is, for a given model $\mathcal{M_I}$ and parameter value $\theta^s$, the model output $\mout=\mathcal{M_I}(\inp,\theta^s)$ was first calculated. Then its mode amplitudes with respect to the PCA were calculated as
\begin{equation}
a_\alpha^{s} = (\mout-\bar{\out}) \cdot \boldsymbol{v}_{\alpha}.
\end{equation}
Lastly, the error term was calculated by adopting the $L_2$ norm as
\begin{equation} 
e^2 =  \| (\mout -\bar{\out}) - \sum_{\alpha=1}^M a_\alpha^s \boldsymbol{v}_{\alpha} \|^2.
\end{equation}
Thus, the likelihood was calculated as
\begin{align}
p(\bayesequal \mid \theta^s,\mathcal{M_I}) =&  \left[ \prod_{\alpha=1}^M \frac{1}{\sqrt{2\pi} s_\alpha} \exp\left(- \frac{(a_\alpha^s)^2}{2s_\alpha^2} \right) \right] \nonumber \\
&  \times\frac{1}{\sqrt{2\pi(\sigma^2+\sigma_n^2)}} \exp\left(- \frac{e^2}{2(\sigma^2+\sigma_n^2)} \right)\label{likelihood-func}.
\end{align}

\subsubsection{Choice of parameter prior distributions}
Choosing the prior probability distribution of parameters for each model, $p(\theta \mid \mathcal{M_I})$, is an important step in the proposed framework. In the absence of any prior information about the parameters, an uninformed prior---specifying a uniform distribution in a range---can be assumed. However, there is no obvious way to choose an appropriate range for each parameter. To make this choice of range consistent across models, the following approach was used. A classical curve-fitting technique was used to fit the model output $\mout$ to the mean response $\bar{\out}$ resulting in best-fit parameter values $\bar{\theta}$ (Table~\ref{fit-params}). Details of the classical fitting procedure and a remark on its relation to the likelihood function are provided in \ref{optim}. Subsequently, a range of $\theta_i \in \left[ {\bar{\theta}_i}/{10}, 10(\bar{\theta}_i+1)  \right]$ for each parameter $\theta_i$ was used (Table~\ref{model-priors2}), thus spanning two orders of magnitude around the best-fit values of the parameters. A different procedure was used for the structural parameter $\kappa$: its distribution was assumed to be uniform in the entire admissible range (usually [0, 1] or [0, 1/3]). The ranges of the parameters used for each model are listed in Table~\ref{model-priors2}.

\subsection{Post-processing and statistical tests}
The simulations performed for computing the Monte Carlo integral in Eq.~\eqref{MC} can also be used to obtain further insights. For example, the posterior distributions of the parameters, Eq.~\eqref{bayes1}, for each model can be computed. From the posterior distributions, point estimates of the expected parameter values can be defined as
\begin{equation}
    \mathbb{E}(\theta) = \int\limits_\theta \theta \, p(\theta \mid \bayesequal ,\mathcal{M_I})  \, \d\theta\label{eq:exp-params}
\end{equation}
and subsequently approximated via Monte Carlo approximation. Another point estimate, the maximum a posteriori (MAP) estimate $\theta^{\text{MAP}}$, was also approximated from the Monte Carlo samples as
\begin{equation}
    \theta^{\text{MAP}} \approx \argmax\limits_{s\in\left\{1,\dots,S\right\}} p(\theta^s \mid \bayesequal ,\mathcal{M_I}).\label{MAP-def}
\end{equation}
Equivalently, the expected model output and its variance,
\begin{align}
    \mathbb{E}(\mout) &= \int\limits_\theta \mout \, p(\theta \mid \bayesequal ,\mathcal{M_I})  \, \d\theta \text{ and} \\
    \mathbb{V}(\mout) &= \int\limits_\theta \left[\mout - \mathbb{E}(\mout)\right]\otimes\left[\mout - \mathbb{E}(\mout)\right] \, p(\theta \mid \bayesequal ,\mathcal{M_I})  \, \d\theta,
\end{align}
were approximated using the Monte Carlo integration. Lastly, to obtain a histogram of the posterior probability distributions of each of the model parameters $\theta_i$, the range of each parameter was divided into 20 equal-sized bins, $B_i^J := \left[\theta_i^J,\theta_i^{J+1}\right]$, $J=1,\dots,20$. Then, the posterior probability distribution $p(\theta\mid\bayesequal,\mathcal{M_I})$ was marginalized with respect to the other parameter $\theta_{j\ne i}$ to obtain following discrete probability:
\begin{equation}
    p\left(\theta_i\in B_i^J\right) \propto \int\limits_\theta H\left(\theta_i,B_i^J\right) p(\theta\mid\bayesequal,\mathcal{M_I})\,\d\theta,
\end{equation}
where 
\begin{equation}
    H\left(x,B\right)=\begin{cases}
    1 & \text{if }x\in B \\
    0 & \text{ otherwise}
    \end{cases}.
\end{equation}
The above integral was also approximated using Monte Carlo.

For finding differences between cusp types, independent samples t-test was used to compare the modal amplitudes. For finding correlations between modal amplitudes and tissue thicknesses, Pearson's correlation coefficient was used.

\section{Results}
\subsection{Data pre-processing}
\begin{figure}[h!]
\centering
\includegraphics[width=0.7\columnwidth]{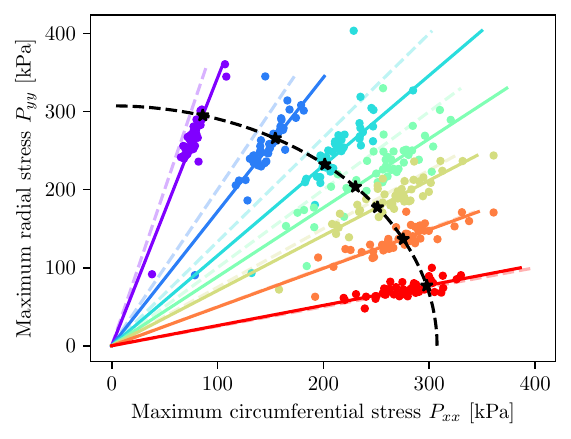}
\caption{The applied stresses (dots) deviated slightly from the target ratio (faint dashed colored lines), where the mean achieved ratio is shown in solid colored lines. The mean magnitude of the maximum applied stress $P^{\mathrm{max}}$ is plotted as a black dashed circular arc\label{force-ratio}. The target stresses for each protocol are the intersection of the circular arc and solid lines, and are denoted with black $*$.}
\end{figure}

The considered ratio $\phi$ varied slightly from the target values (Fig.~\ref{force-ratio}). The mean values of the applied ratios were $\phi_r \in \left\{ 0.29, 0.58, 0.87, 1.13, 1.42, 2.02, 3.87 \right\}$, and the mean magnitude of the maximum applied stress was $P^{\text{max}}=307.4$ kPa. These mean values of stress ratios $\phi$ and maximum stress magnitude $P^{\text{max}}$ were used for evaluating the common input vector $\inp$ (representing the target applied stresses, Eq.~\ref{input}) and thereafter compute the model outputs $\out$ (representing the resulting stresses and stretches, Eq.~\ref{output}).  

\begin{figure*}[h!]
\subfigimg[width=\textwidth]{a)}{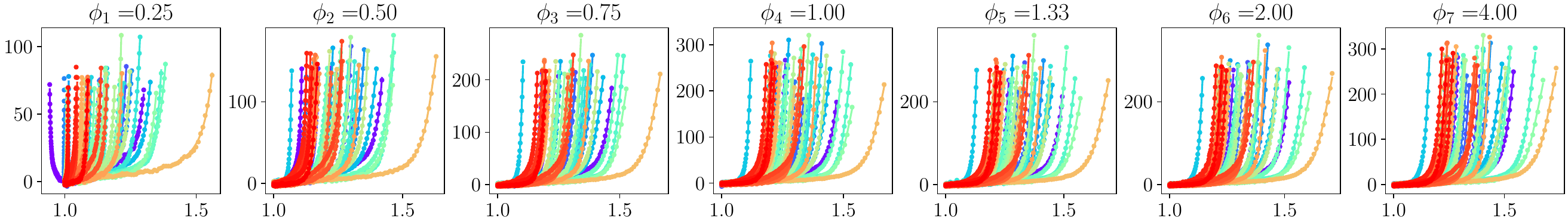}\\
\subfigimg[width=\textwidth]{b)}{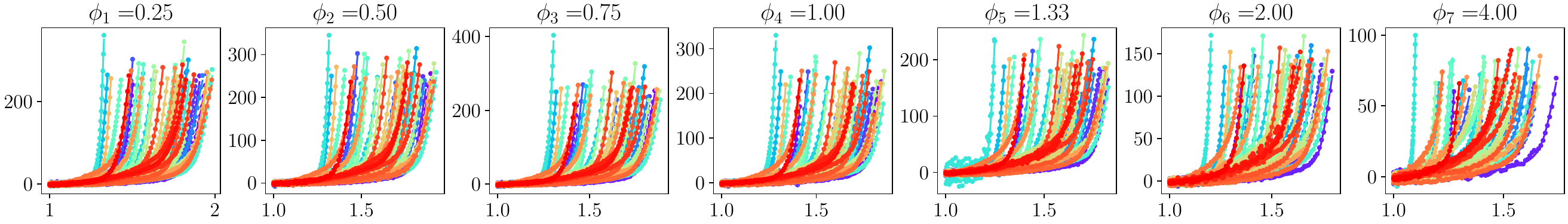}
\caption{The experimental data for $N=46$ samples (dots) and the fitted interpolating function (line) in the (a) fiber and (b) cross-fiber direction; horizontal axes are stretches and vertical axes are stresses in [kPa]. Each color represents a different tissue sample of total $N=46$ specimens. \label{regressed-data}}
\end{figure*}

\begin{figure*}[h!]
\subfigimg[width=\textwidth]{a)}{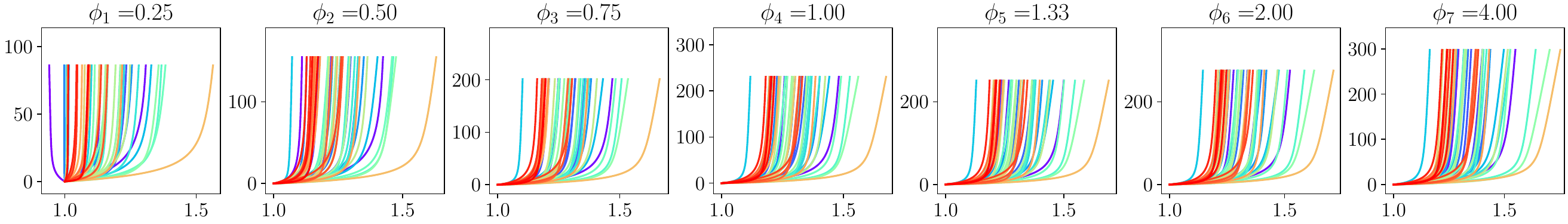}\\
\subfigimg[width=\textwidth]{b)}{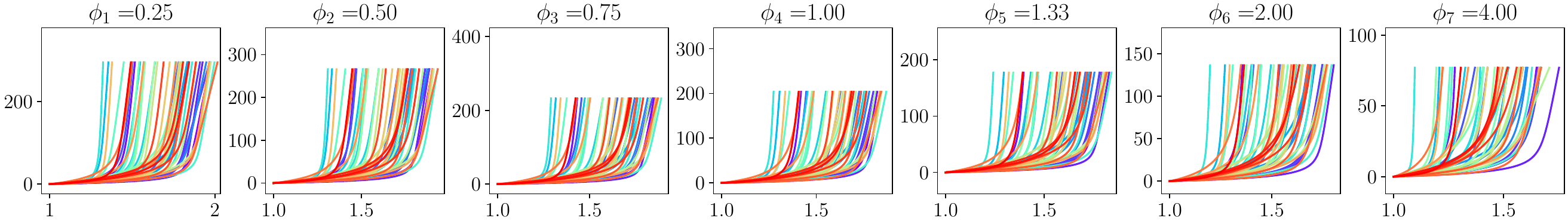}
\caption{The interpolated data in the (a) fiber and (b) cross-fiber direction; horizontal axes are stretches and vertical axes are stresses in [kPa]. Each color represents a different tissue sample of total $N=46$ specimens. \label{interp-data}}
\end{figure*}

The stress-stretch curves at all ratios for all 46 samples are plotted as points in Fig.~\ref{regressed-data}. The 1D model (Eq.~\ref{freed-func}) fit all the stress-stretch curves well (solid lines in Fig.~\ref{regressed-data}), without causing any issues of overfitting, oscillations, or ill-conditioning. The coefficient of determination of the fit was $R^2>0.927$ for all curves, with the mean value being $\left< R^2 \right> = 0.996$. The fitted values of parameters and coefficients of determination are provided as Supplementary Information (SI). Thus, the function also allowed reliable interpolation/extrapolation. The interpolated stress-stretch curves for the common input vector $\inp$ are shown in Fig.~\ref{interp-data}. These curves were used as the input to the proposed model selection framework, starting with the principal component analysis.

\subsection{Principal component analysis}

\begin{figure*}[h!]
\subfigimg[width=\textwidth]{a)}{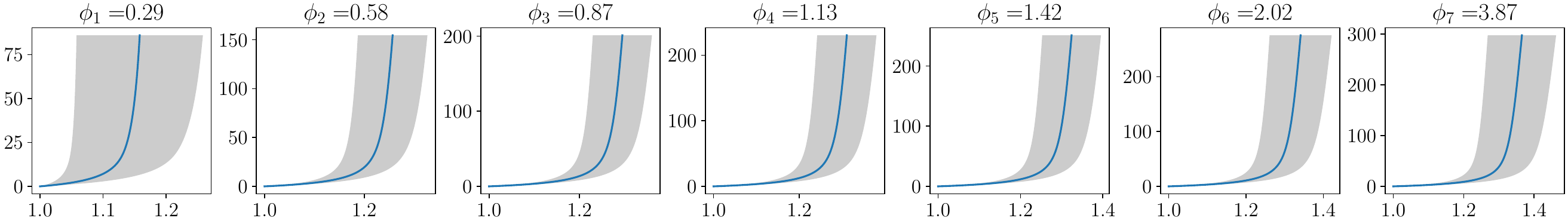}\\
\subfigimg[width=\textwidth]{b)}{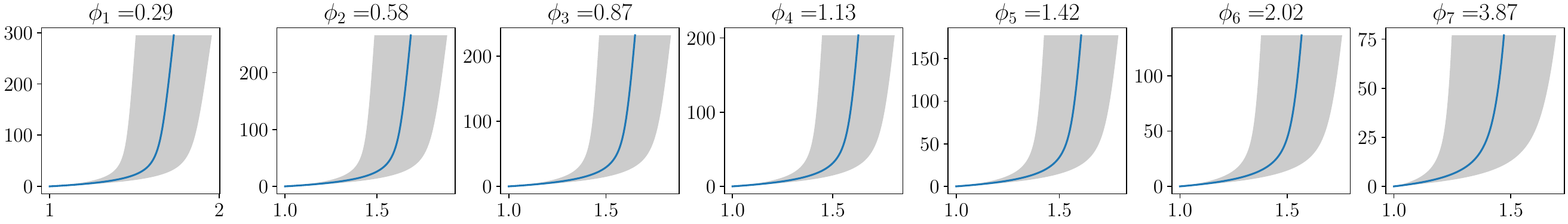}
\caption{The mean stress-stretch response (blue lines) in the (a) fiber and (b) cross-fiber direction; horizontal axes are stretches and vertical axes are stresses in [kPa]. The shaded gray area denotes one standard deviation.\label{mean-response}}
\end{figure*}

\begin{figure}[h!]
\centering
\includegraphics[width=0.6\columnwidth]{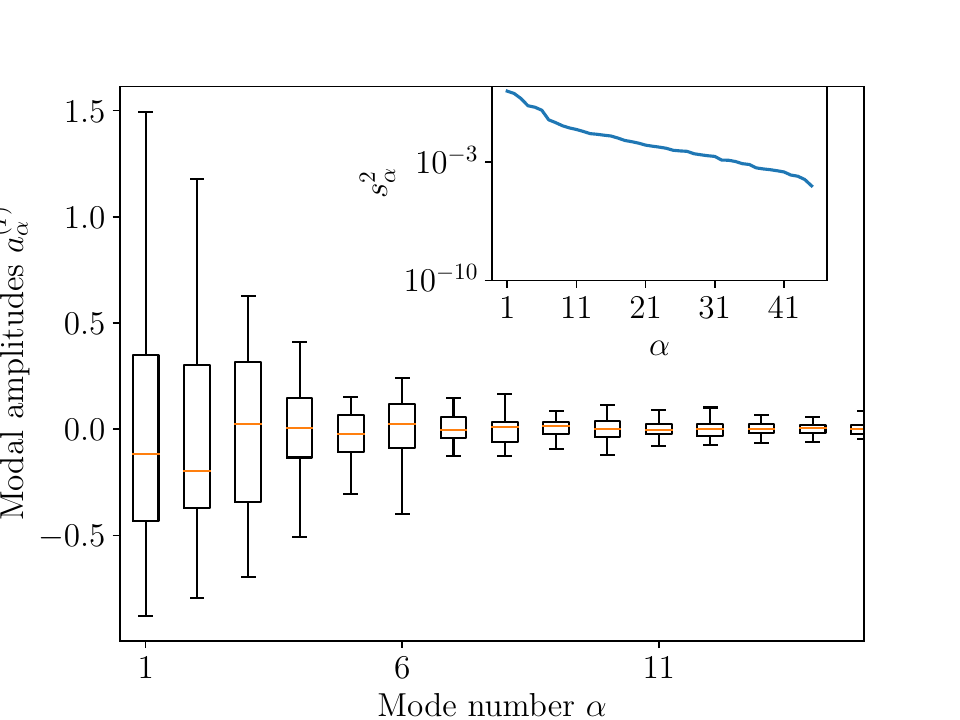}
\caption{A boxplot of the amplitudes of PCA modes in the experimental data, from which a normal distribution is constructed with mean zero and variance equal to $s_\alpha^2$ (inset).\label{mode-amps}}
\end{figure}

\begin{figure*}[h!]
\subfigimg[width=\textwidth]{a)}{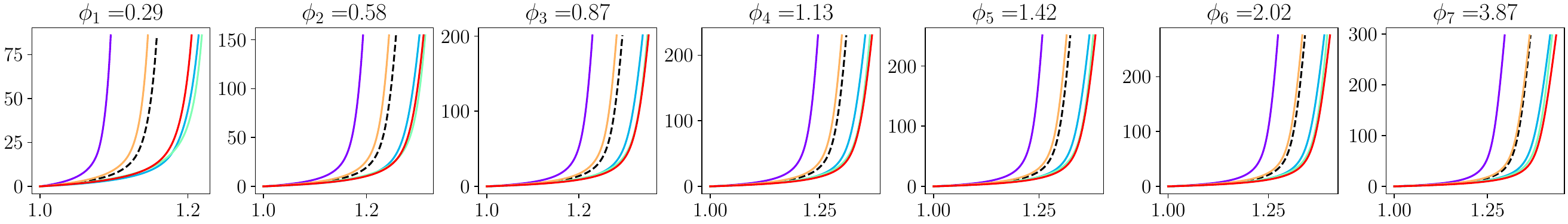}\\
\subfigimg[width=\textwidth]{b)}{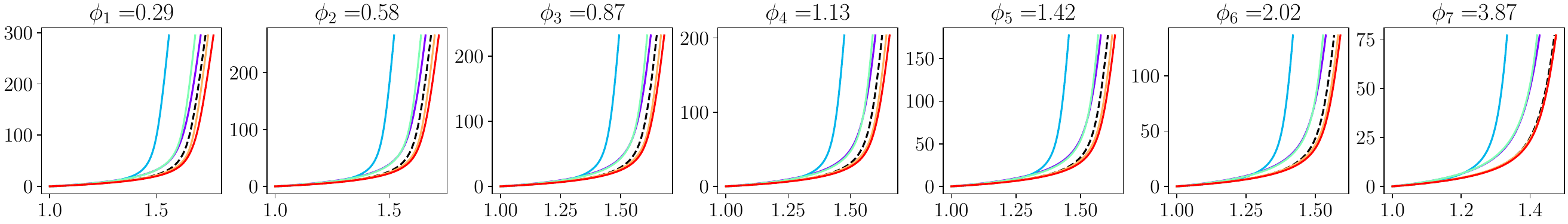}\\
\includegraphics[width=\textwidth]{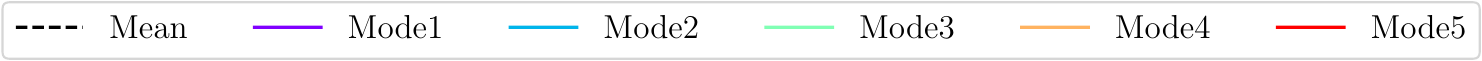}
\caption{First five principal modes of stress-stretch response in the (a) fiber and (b) cross-fiber direction; horizontal axes are stretches and vertical axes are stresses in [kPa]. For an animation of the modes, see SI. \label{eigvecs}}
\end{figure*}

The mean stress-stretch response is shown in Fig.~\ref{mean-response}, with the variation of one standard deviation depicted as shaded area. The amplitudes of all the normal modes calculated using PCA are shown in Fig.~\ref{mode-amps} (see SI for an animation of the first five PCA modes). The main boxplot shows the variation in the modal amplitudes of the experimental data, which is only dominant for the first five modes. The inset in Fig.~\ref{mode-amps} shows the singular values of each mode on a log-scale, which decrease exponentially. The first five dominant PCA modes are plotted in Fig.~\ref{eigvecs}.

\begin{figure*}[h!]
\subfigimg[width=\textwidth]{a)}{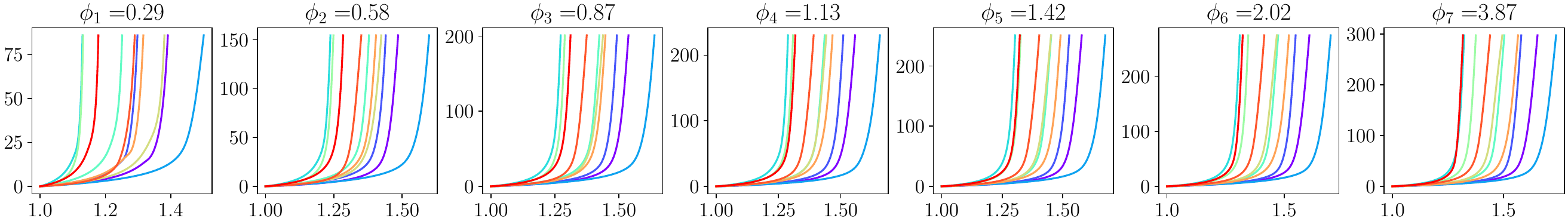}\\
\subfigimg[width=\textwidth]{b)}{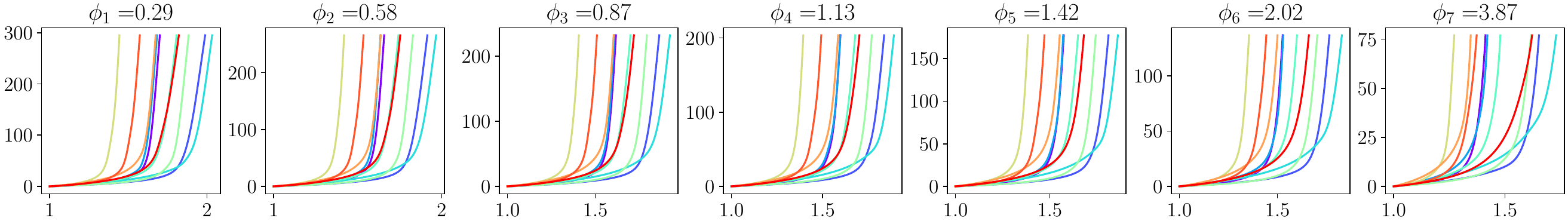}
\caption{Ten synthetic samples' stress-stretch response using Eq.~\eqref{statistical-model} in the (a) fiber and (b) cross-fiber direction; horizontal axes are stretches and vertical axes are stresses in [kPa]. Each color represents one synthetic sample.\label{aritificial-data}}
\end{figure*}

Based on the PCA, a statistical model (Eq.~\ref{statistical-model}) was constructed. The framework allows generation of synthetic dataset based on this statistical model. Ten random samples are shown in Fig.~\ref{aritificial-data}, which demonstrates that the statistical model captures the variation in the actual dataset.

\subsection{Model probabilities}
\begin{figure}[h!]
\centering
\includegraphics[width=0.7\columnwidth]{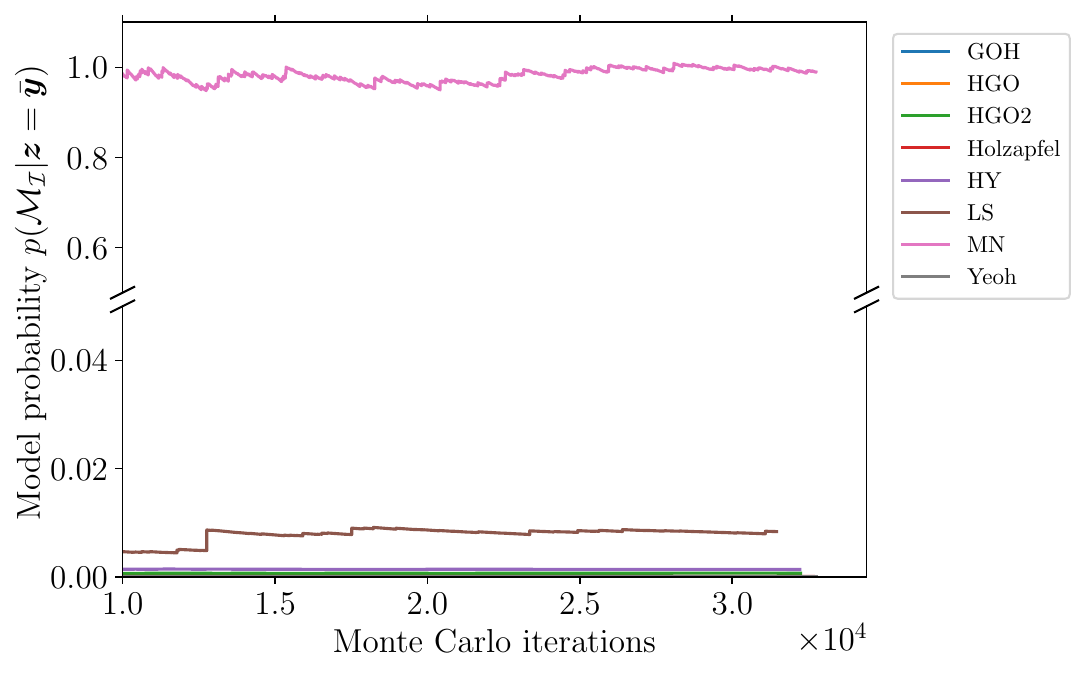}
\caption{Relative probabilities of the eight hyperelastic models calculated using the proposed framework with $M=11$ versus number of Monte Carlo iterations.
\label{cal-probs12}}
\end{figure}

\begin{figure}[h!]
\centering
\subfigimg[width=0.49\columnwidth]{a)}{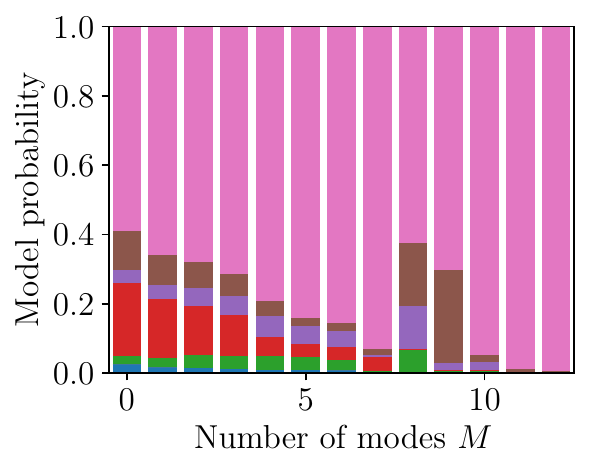} 
\subfigimg[width=0.49\columnwidth]{b)}{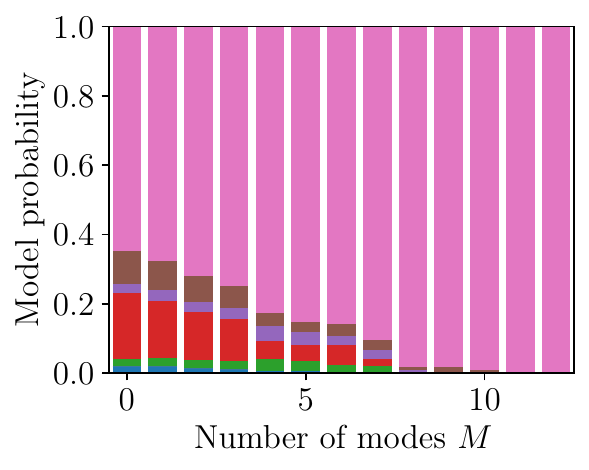} \\
\subfigimg[width=0.49\columnwidth]{c)}{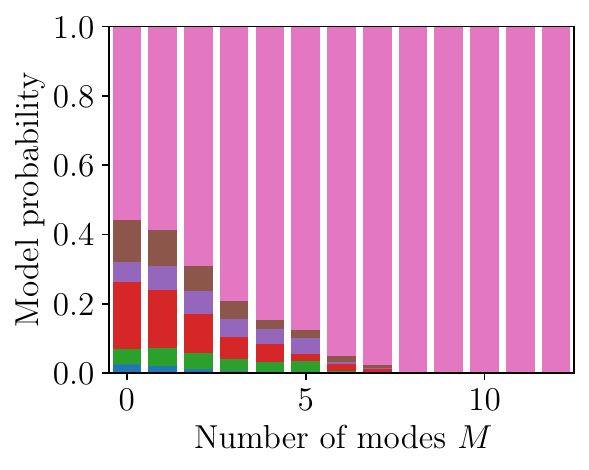} 
\subfigimg[width=0.49\columnwidth]{d)}{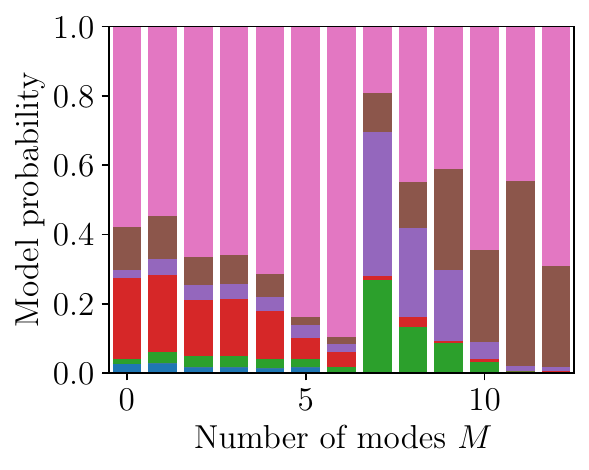} \\
\includegraphics[width=0.55\columnwidth]{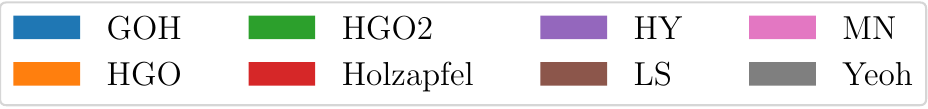}
\caption{The probabilities of the eight hyperelastic models versus number of PCA modes retained $M$ for (a) all, (b) LCC, (c) RCC, and (d) NCC types. \label{cal-probs22}}
\end{figure}

Using the proposed framework, the convergence of Monte Carlo integration was confirmed by plotting the model probabilities for $M=11$ versus number of iterations (Fig.~\ref{cal-probs12}). Clearly, all the computations were converged. The most probable model comes out to be the May--Newman model, followed by the Lee--Sacks model. If the number of modes $M$ retained in the statistical model are varied, the model probabilities vary, as shown in Fig.~\ref{cal-probs22}a. Interestingly, if no principal modes are considered ($M=0$), i.e., the comparison of classical fitting to the mean is used, five out of the eight models have roughly similar probabilities, with May--Newman model being the most probable. This means that the five of the models are able to describe the mean response well. However, as variations along the principal modes are included in the statistical model, it becomes possible to differentiate the models. For $M\ge11$, the May--Newman model performs significantly better that the other models. 

If we categorize the samples by cusp types (LCC, RCC, and NCC), the relative probabilities of the eight models for each cusp type show varying behavior (Fig.~\ref{cal-probs22}b--d). For both LCC and RCC tissues, the May--Newman model still has the largest probability, and there is a clear convergence of model probabilities as modes are increased. In contrast, for NCC tissues, the probabilities oscillate as we increase the number of modes considered. At $M\ge11$ the probabilities of Lee--Sacks and May--Newman models are comparable. This indicates that both May--Newman and Lee--Sacks models are equally good at describing the data of NCC tissues, and one cannot be ruled out in favor of the other. Moreover, some of the other models, e.g., Humphrey--Yin model, perform well for certain values of $M$ and should not be discarded. Generally, the NCC tissues show a unique behavior at mode $M=7$ that affects all model probabilities and will require further investigation in the future. 

\subsection{Expected and MAP parameter values}

Since the MN model clearly outperforms other models based on the proposed Bayesian framework, we analyze its properties further. Using the Monte Carlo integration, for MN model, the expected values of its parameters (Eq.~\ref{eq:exp-params}) were calculated to be $\mu=11.26$ kPa, $k_1=305.66$ kPa, $k_2=1.74$, and $k_3=299.73$. 
For comparison, the MAP estimates of its parameters (Eq.~\ref{MAP-def}) were found to be $\mu=13.74$ kPa, $k_1=253.77$ kPa, $k_2=1.1$, and $k_3=335.72$. When compared with the parameter values using classical fit (Table~\ref{fit-params}), a large difference is noticed. This is due to a fundamental difference in the two approaches. While, the classical fit tries to match the model response to the mean response, the presented approach prefers to match the \emph{shape} (i.e. principal modes of variation from the mean) of stress-stretch curves from a model to those observed in the experiments. 

\begin{figure*}[h!]
\centering
\subfigimg[width=\textwidth]{a)}{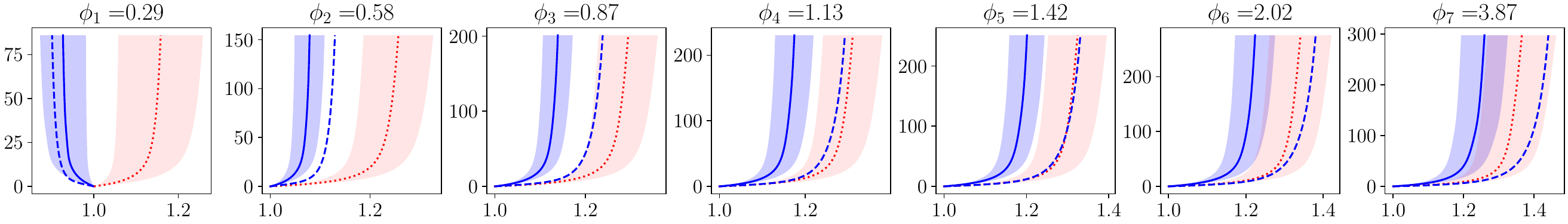}\\
\subfigimg[width=\textwidth]{b)}{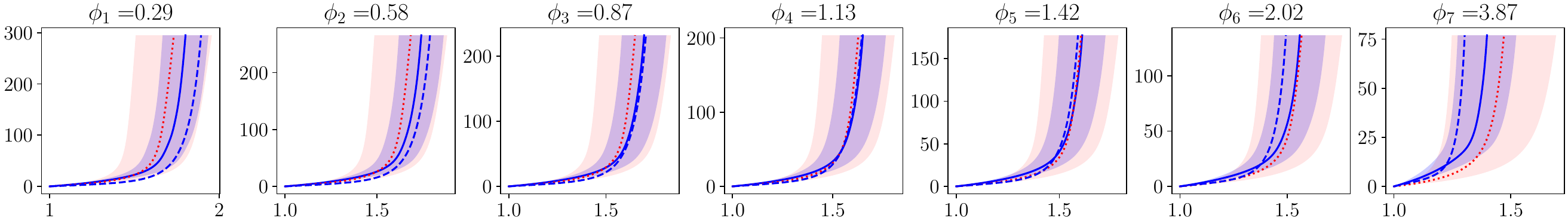}\\
\caption{Posterior response using May--Newmann model with mean (solid line) and variation (shaded region) in blue color compared to the data in red color, in the (a) fiber and (b) cross-fiber direction. For comparison, the classic fit of the MN model is plotted with dashed blue lines. Horizontal axes are stretches and vertical axes are stresses in [kPa].\label{posterior-plot}}
\end{figure*}

\begin{figure}[h!]
\centering
\includegraphics[width=0.9\textwidth]{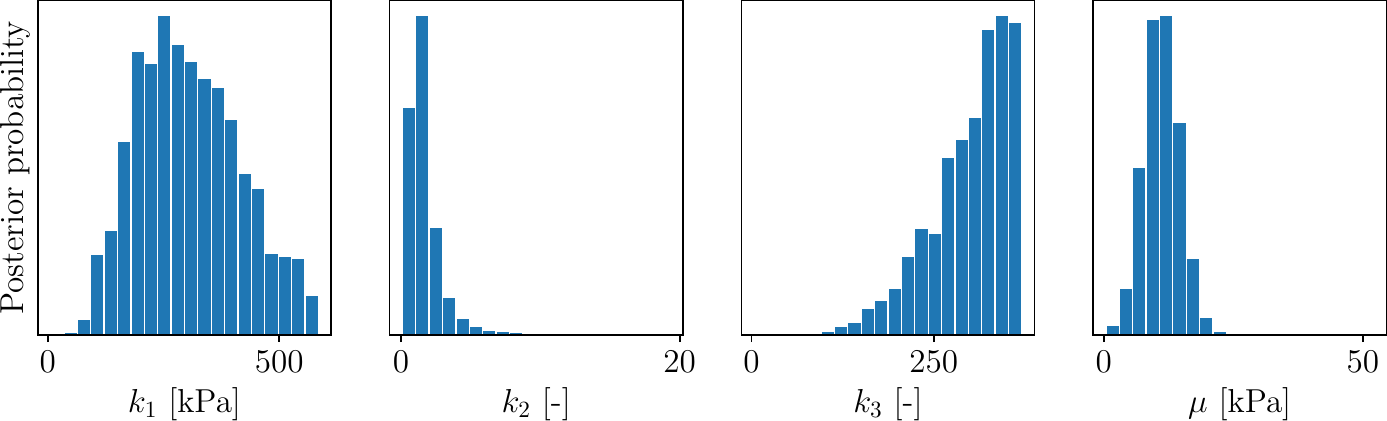}
\caption{Histogram of the posterior distribution of parameters of the May--Newman model.\label{posterior-plot2}}
\end{figure}

The resulting stress-stretch curves and their variations were computed for the MN model and are shown in Fig.~\ref{posterior-plot}. While the MN model captures the cross-fiber response well, its response along the fibers deviates from the data, especially for low $\phi$. This issue is also present in all other models considered (results not shown for brevity), and it is related to the complex coupling of fiber and matrix. This mismatch between the data and the considered models indicates that none of the considered models are perfectly suited for aortic valve tissues and signifies the need for continued developments in the field of constitutive modeling. The difficulty in selecting a model in the absence of a clear, perfect fit highlights the need for a systematic framework that allows an objective and easy-to-interpret comparison of models. Lastly, the posterior distributions of the MN model parameters are shown in Fig.~\ref{posterior-plot2}, which could be used for population-based studies in the future.

\section{Discussion}
The coupling between fiber and cross-fiber directions in soft tissues produces a complex behavior, which is only captured when multiple biaxial loading ratios are used in a biaxial mechanical testing setup. Many constitutive models have been developed to describe this observed behavior and provide better predictive capabilities. However, commonly these models are fit only to the mean response, effectively regarding the variability observed in biological samples as being random. In contrast, the PCA results highlight that the observed variation is not random and should be considered when matching with a model's response.

\subsection{Efficacy of the proposed framework}
In the present study, we proposed a novel framework that accounts for the inter-sample variability and allows for the computation of the probabilities of chosen constitutive models. Notably, instead of finding the best fit as is commonly done in the literature, the proposed framework aims to find the model that can capture \emph{the variation} seen in the experimental dataset. This approach depends on having data from a large number of tissue samples. The large and, potentially variable, number of data points for each sample pose a critical challenge in the application of statistical tools. Herein, one-dimensional interpolation and PCA-based techniques were used to tackle these challenges, and were found to be effective. 

For the interpolation, it was not trivial to establish a function that can fit all the observed stress-stretch curves. The one-dimensional stress-strain function proposed by Freed and Rajagopal \cite{freed2016promising} worked extremely well (Fig.~\ref{regressed-data}). Since the function is based on elasticity theory, by construction, it excluded non-physical responses, such as oscillations that are commonly observed in the low-stress regime of the biaxial stretching data. Moreover, even with just three parameters, it was able to fit all of the $2N\times R=644$ stress-stretch curves ($R^2>0.927$ and mean $\left<R^2\right>=0.996$). 

In addition, PCA was used to reduce the dimensions of the dataset and thereby establishing a computationally useful statistical model. Since the PCA modes are orthogonal, the amplitudes of the PCA modes are independent, by construction. Thus, PCA reduces the problem to constructing distributions of several scalar variables. In this work, each modal amplitude was assumed to have a normal distribution. However, this assumption can be relaxed if the hypothesis of normality can be rejected with sufficient number of samples.

Once the statistical model is established, the proposed framework is straightforward to implement, and the Monte Carlo integration scheme is trivial to parallelize. The results converged for all Monte Carlo simulations, with typical computational times of 30 minutes with a 16-core CPU. Thus, the framework is computationally feasible while providing descriptive statistical insight. The fact that the framework was able to distinguish between similar models (all of them dependent on $I_1$ and $I_4$ with exponential terms) and pick up the differences between leaflet types, also observed in the PCA, is remarkable and substantiates its reliability.

Lastly, the Bayesian framework offers some practical advantages compared to the classical parameter-fitting approach. While in-general having a higher number of parameters gives a model an advantage in fitting the data better, this advantage is naturally taken into account in the Bayesian setting wherein models with more parameters are penalized \cite{doi:10.1142/S0218202513500103}. Moreover, finding a unique global minima in parameter-fitting can be challenging for problems with either insufficient data to differentiate the parameters or with highly (or perfectly) correlated parameters \cite{bioengineering6040100,aggarwal-exponential}. The proposed framework circumvents these issues by integrating over the parameter space (see SI for a simple demonstration of these features). Nevertheless, finding a suitable model is a multi-faceted problem, and if the uniqueness of parameters is of interest, that aspect could be accounted for in choosing a model. 

\subsection{Insights into tissue mechanics}
Our results highlight several important characteristics of soft tissues, in general, and for aortic valve cusps, in particular. The observed variation in the stress-stretch response of tissues was larger in the fiber direction compared to the cross-fiber direction (Fig.~\ref{mean-response}). Moreover, it is clear that even the most probable model (i.e., the May--Newman model) does not capture the stress-stretch curves at all of the biaxial loading ratios (Fig.~\ref{posterior-plot}). This means that there is a trade-off while matching the models to the measurements. Nonetheless, the proposed framework naturally accounts for this trade-off by integrating over the parameter space, weighted appropriately.

From the considered eight hyperelastic models and based on the information available through the utilized experimental dataset, the most probable model for AV tissues was the one proposed by May--Newman \cite{may1998constitutive}. Although this model was originally proposed for mitral valve tissues, the same model form has been adopted for the aortic valve tissue as well \cite{10.1115/1.3127261}. The second most probable model was the one by Lee and Sacks \cite{LEE20142055}, which has been used for mitral valve, tricuspid valve and bioprosthetic valve. 

\begin{figure}[h!]
\centering
\includegraphics[width=0.5\columnwidth]{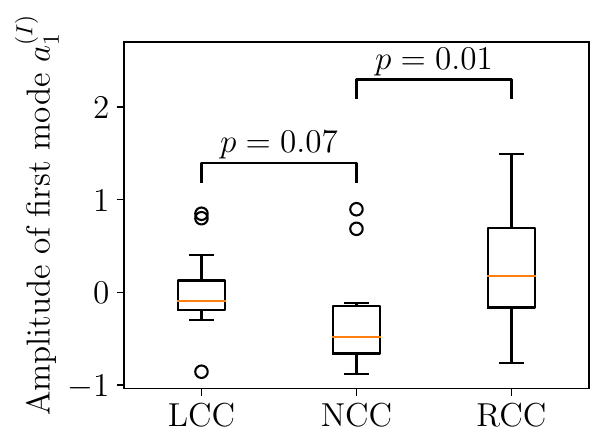}
\caption{Independent samples $t$-tests between the modal amplitudes of different cusp types showed a significant difference in the second principal mode between the NCC ($n=16$) and the other two cusp types ($n=15$ each), which is consistent with the finding that the probability of models is different for the NCC.\label{mode1-stat}}
\end{figure}

A key finding from this study is that different models are suitable for different AV cusp types. Specifically, the Lee--Sacks model \cite{LEE20142055} and the May--Newman model \cite{may1998constitutive} were equally probable for the NCC, while the May--Newman model was the most probable model for the LCC and RCC. In order to further investigate these differences, the amplitudes of the PCA modes for the three cusp types were compared, and noticeable differences were found in the first mode amplitudes (Fig.~\ref{mode1-stat}). Specifically, there was a statistically significant difference between the NCC and RCC samples ($p<0.05)$. Although the difference between the NCC and LCC samples was not established to be statistically significant ($p=0.07$), it could be due to the limited sample size in this study. The different stress-stretch behavior of the NCC might also be related to its different physiology (no coronary flow) and the observed differences in its geometry relative to the other two cusp types \cite{https://doi.org/10.1002/ca.10149}.

Furthermore, when processing the data from biaxial mechanical testing, it is common to work with stresses, essentially factoring out tissue thickness. However, for valves, their thickness is an important design feature. To investigate the relation between the thickness and stress-strain response of tissues, the classical Pearson's correlation coefficient was calculated between the modal amplitudes and tissue thicknesses. The second and third modes showed a statistically significant correlation (Table~\ref{correlations}). This indicates a correlation between the tissue thickness and its stress-strain behavior, which should be accounted for while constructing population-level models. One way of achieving this is by working with membrane tension rather than stress, thereby incorporating the tissue thickness into the model parameters. 

\begin{table}[h!]
\centering
\begin{tabular}{lcc}
\hline
\bf{Mode} & \bf{Correlation coefficient $\boldsymbol{r}$} & \bf{p-value} \\
\hline
0 & $-0.25$ & 0.09 \\
1 & $+0.36$  & 0.02 \\
2 & $-0.33$ & 0.03 \\
3 & $+0.12$  & 0.44 \\
4 & $-0.22$ & 0.14 \\
\end{tabular}
\caption{Pearson's correlation coefficient between tissue thickness and amplitudes of the first five PCA modes and corresponding p-values}
\label{correlations}
\end{table}

\subsection{Limitations}
While the proposed framework allows for an \emph{objective} comparison between constitutive models, the objectivity should be interpreted within the scope of the problem---i.e. within the scope of the experimental data used and the models considered in the study---and from a Bayesian perspective. This means that the results are only valid for the data used and subject to the prior probabilities used. Having more experimental data (e.g., shear deformations have been shown to be important for constitutive modeling \cite{doi:10.1177/1081286507084411}) would make the results more reliable. Similarly, microstructural information on the tissues, if available, could be used to either fix or more tightly constrain the fiber dispersion parameters (such as $\kappa$) in the models.

The list of models considered is also limited by practical limitations. While, ideally, one would include include all the models available in the literature, this process is practically infeasible. Therefore, only the most widely used constitutive models for soft tissues were used in this study. However, there could be other models in the literature that might be more suitable, such as those with nonlinear contributions from the isotropic matrix \cite{10.1115/1.4037916}, those with $I_2$ and $I_5$ invariants \cite{ANSSARIBENAM2021103486, destrade2013least, MURPHY201390}, those with logarithmic functions in the strain energy density \cite{horgan2003description, doi:10.1177/108128028477}, or meso-scale or multi-scale models. Thus, the framework only compares chosen models, and any model not included in this framework cannot be excluded or disregarded.

Results from the proposed framework depend on the choice of prior parameter distributions. Practically, choosing the parameters' prior distributions in a manner that is consistent across models is not trivial. Too wide of a distribution indicates uncertainty, which may lower the final model probability. Equivalently, a narrow distribution may limit the model's capability to cover the observed spread in the data. One approach for choosing consistent parameter ranges was used in this study, however other approaches should be tested in the future. To confirm the results in this study, a uniform prior in the range obtained by fitting a model to each of the individual samples was also tested. The resulting model probabilities were largely the same (results presented in SI), further adding confidence to our results. 

Another limitation of the proposed framework is its reliance on having a large enough sample size. Thus, more data would allow construction of better statistical models and higher confidence in the results. These limitations will be addressed in the future work in this direction, that is outlined next.

\subsection{Future work and conclusion}
In the future, the proposed framework will be applied to other tissue types from animals and humans. Additional models that are known to satisfy continuum and thermodynamic requirements will be included in the investigation, allowing a comparison of wider range of models. Specifically, models based on other invariants \cite{ANSSARIBENAM2021103486, destrade2013least, MURPHY201390} and other functional forms \cite{horgan2003description, doi:10.1177/108128028477} will be studied, as well as structural and multi-layer models that are specifically developed to describe the coupling. While an equal prior probability was assigned to all the models considered here, any concerns regarding the stability/convexity/thermodynamic requirements could be reflected in a reduced prior probability. This study focused on elastic behavior under quasi-static loading, but a similar framework could be developed to compare models that describe the viscoelastic behavior at varying loading rates observed in valve tissues \cite{doi:10.1098/rsos.160585}.

In this study, only the biomechanical parameters were considered to be random, while the stress-free state of the tissues were assumed to be known. However, the stress-free (i.e., reference) state of tissues is well-known to be difficult to assess \cite{AGGARWAL20162481,LAURENCE2022}. In valve tissues, this is due to the existing pre-stresses at different scales \cite{10.1115/1.2768111} and long toe regions in the stress-strain response. A novel feature of the proposed framework is that it can account for uncertainty in the reference state by considering prestresses to be random variables. This extension will be undertaken in the future. 

In conclusion, the framework will facilitate an objective comparison of constitutive models against experimental data for different tissues. In conjunction with optimal design of biaxial experiments \cite{axioms2021} and improved parameter estimation techniques \cite{bioengineering6040100,aggarwal-exponential},  work in the proposed direction will lead to the development of more predictive biomechanical models that are representative of population, not just individual patients.

\section*{Acknowledgments}
We thank Prof Nele Famaey and Dr Heleen Fehervary for insightful discussions. This work was supported by grant EP/P018912/2 from the Engineering and Physical Sciences Research Council of the UK and grant R01 HL159475 from the National Institutes of Health.


\section*{Conflicts of interest}
The authors have no relevant financial or non-financial interests to disclose.

\appendix

\section{Bayes' Theorem}\label{bayes-notation}

For two continuous random variables $A$ and $B$, let the joint prior probability density function be denoted by $p(A,B)$. Further, the prior marginal probability densities of $A$ and $B$ are denoted as $p(A)$ and $p(B)$, respectively. The posterior probability density of $A$ given $B$ (known as the conditional probability) is denoted as $p(A\mid B)$ and is given by the Bayes' theorem:
\begin{equation*}
    p(A \mid B) = \frac{p(B \mid A) p(A)}{p(B)},
\end{equation*}
where $p(B\mid A)$ is the likelihood term. The denominator on the right hand side is also the normalization term as shown below:
\begin{equation*}
    p(B) = \int p(A, B) \, \mathrm{d}A = \int p(B \mid A) p(A) \, \mathrm{d}A.
\end{equation*}

\section{Classical Fitting}\label{optim}
Given the mean output $\bar{\out}$, the classical way of fitting a model is to find the model parameters such that the model output $\mout(\theta)=\mathcal{M_I}(\inp,\theta)$ is closest to the mean output $\bar{\out}$ in some norm. A commonly used  $L_2$ norm is adopted here in the present study, i.e., the sum of squares of each component difference. Mathematically, we write:
\begin{equation*}
   \bar{\theta} := \argmin\limits_{\theta} \| \mout(\theta) - \bar{\out} \|^2.
\end{equation*}
The above minimization was performed using the trust-region reflective algorithm implemented in SciPy. 

When we do not consider any of the PCA modes (i.e., $M=0$) and ascribe all variation in the data as error, the likelihood function \eqref{likelihood-func} becomes
\begin{equation*}
    p(\bayesequal \mid \theta^s,\mathcal{M_I}) 
    = \frac{1}{\sqrt{2\pi(\sigma^2+\sigma_n^2)}} \exp\left(- \frac{e^2}{2(\sigma^2+\sigma_n^2)} \right),
\end{equation*}
where the error term also simplifies to $e^2 = \|\mout - \bar{\out} \|^2$. It is easy to see that the maximum likelihood happens when $e^2$ is minimized. In other words, the fitted parameters  $\bar{\theta}$ also correspond to the maximum likelihood with $M=0$. In contrast, as the PCA modes are included, the likelihood depends not only on the error term (i.e., random variations), but also on the specific variations along the included PCA modes.

\bibliographystyle{elsarticle-num-names} 
\bibliography{refs,reviews/new-refs} 

\end{document}